\documentclass[aps, prd, onecolumn, tightenlines, notitlepage, superscriptaddress, nofootinbib, preprintnumbers, floatfix,showkeys,11pt,altaffilletter]{revtex4-2}

\usepackage{upgreek}

\usepackage[official]{eurosym}
\usepackage[normalem]{ulem}
\usepackage{amstext}
\usepackage{graphicx}
\graphicspath{{figures/}}
\usepackage{url}
\usepackage{color}
\usepackage{ulem}
\usepackage[version=4]{mhchem}
\usepackage[utf8]{inputenc}
\usepackage{fontawesome}
\usepackage{yfonts}

\pdfoutput=1
\usepackage{textcomp}
\usepackage{comment}
\usepackage{yfonts}
\usepackage{epsfig,amsfonts,mathrsfs,graphicx,color,slashed,multirow}
\usepackage{latexsym,graphicx,slashed,color,enumerate,url,cancel,gensymb}
\usepackage{textcomp}

\usepackage[x11names]{xcolor}
\usepackage[colorlinks,pdfstartview=FitV,breaklinks=true]{hyperref}
\usepackage{booktabs}
\usepackage{adjustbox}
\usepackage{textgreek} 
\usepackage{caption, subcaption} 
\usepackage{booktabs} 
\usepackage{float}

\usepackage{lmodern}
\usepackage{ae,aecompl}
\usepackage{appendix}


\definecolor{vdrgreen}{rgb}{0.0, 0.6, 0.0}

\definecolor{persiangreen}{rgb}{0.0, 0.65, 0.58}
\definecolor{mediumpersianblue}{rgb}{0.0, 0.4, 0.65}

\makeatletter
    \newcommand{\colorboxed}[3][white]{\fcolorbox{#2}{#1}{\m@th$\displaystyle#3$}}
\makeatother

\AtBeginDocument{\hypersetup{citecolor=mediumpersianblue,linkcolor=mediumpersianblue,urlcolor=mediumpersianblue}}
\usepackage{appendix}

\begin{document}

\title{{\LARGE Light vector mediators at direct detection experiments}}

\author{Valentina De Romeri}
\email{deromeri@ific.uv.es}
\affiliation{Instituto de F\'{i}sica Corpuscular (CSIC-Universitat de Val\`{e}ncia), Parc Cient\'ific UV C/ Catedr\'atico Jos\'e Beltr\'an, 2 E-46980 Paterna (Valencia) - Spain}

\author{Dimitrios K. Papoulias}\email{dkpapoulias@phys.uoa.gr}
\affiliation{Department of Physics, National and Kapodistrian University
of Athens, Zografou Campus GR-15772 Athens, Greece}

\author{Christoph A. Ternes}
\email{christoph.ternes@lngs.infn.it}
\affiliation{Istituto Nazionale di Fisica Nucleare (INFN), Laboratori Nazionali del Gran Sasso, 67100 Assergi, L’Aquila (AQ), Italy
}


\begin{abstract}
Solar neutrinos induce elastic neutrino-electron scattering in dark matter direct detection experiments, resulting in detectable event rates at current facilities. We analyze recent data from the XENONnT, LUX-ZEPLIN, and PandaX-4T experiments and we derive stringent constraints on several $U(1)'$ extensions of the Standard Model, accommodating new neutrino-electron interactions. We provide bounds on the relevant coupling and mass of light vector mediators for a variety of models, including the anomaly-free $B-L$ model, lepton flavor-dependent interactions like $L_\alpha - L_{\beta}$, $B-2L_e - L_{\mu, \tau}$, $B-3L_\alpha$, and $B+2L_\mu +2 L_{\tau}$ models. We compare our results with other limits obtained in the literature from both terrestrial and astrophysical experiments. Finally, we present forecasts for improving current bounds with a future experiment like DARWIN.
\end{abstract}
\maketitle

\tableofcontents

\section{Introduction}

Dark Matter (DM) Direct Detection (DD) experiments~\cite{Schumann:2019eaa,Baxter:2021pqo}, although primarily designed for DM searches, have been recently recognized as favorable facilities to probe new physics beyond the weakly interacting massive particles (WIMPs) paradigm~\cite{Billard:2021uyg}.
Indeed, and despite the so far eluding WIMP detection,
the latest generation of DM DD experiments --- including low-threshold ($E_\mathrm{thr} = 1$~keV) dual-phase liquid xenon (LXe) detectors developed by the  XENON (Italy)~\cite{XENON:2017lvq,XENON:2018voc,XENON:2022ltv}, LUX-ZEPLIN (LZ) in the USA~\cite{LZ:2015kxe,LZ:2019sgr,LZ:2022lsv}, and PandaX  (China)~\cite{PandaX:2014mem,PandaX:2022ood,PandaX:2024oxq} collaborations ---  have reached world-leading sensitivities also on alternative physics beyond the SM (BSM), for instance  boosted-DM scenarios~\cite{Bringmann:2018cvk,DeRomeri:2023ytt}, dark photons~\cite{Bloch:2020uzh}, axion-like particles~\cite{Ferreira:2022egk}, novel hidden sectors~\cite{An:2014twa}, fermionic absorption DM~\cite{Dror:2019dib} and gravitational waves~\cite{XENONCollaborationSS:2023zbo}.
The physics program of DD facilities will continue and intensify with next-generation detectors~\cite{Billard:2021uyg, Essig:2022dfa, Akerib:2022ort} such as those planned by the DARWIN Collaboration~\cite{DARWIN:2020bnc} that aspire to achieve very low sensitivities leveraging its multi-ton exposure, low background and low threshold. Towards this purpose the XLZD Consortium (XENON, LZ, DARWIN)  aims to develop a 60-ton LXe detector in the next decade~\cite{Aalbers:2022dzr} that will probe WIMPs down to the neutrino fog. At the same time, further advances based on liquid argon (LAr) technology are currently under development by the  Global Argon Dark Matter
Collaboration (GADMC)~\cite{Pesudo:2021nrz}, with  near-term activities focusing on the construction of the DarkSide-20k detector at Laboratori Nazionali del Gran Sasso (LNGS), and the ultimate goal being the construction of a 300-ton argon detector. 

At present, the large fiducial volumes of LXe detectors available at the XENONnT (5.9 ton), LZ (5.5 ton) and PandaX-4T (3.7 ton) experiments have already allowed the collaborations to accumulate exposures of 1.16, 0.90, and 0.63 ton $\times$ year, respectively, in a rather short amount of data-collection time during their first run. 
This, together with their extremely low-threshold capabilities\footnote{Note that the energy threshold is different depending on the detector materials. In the case of liquid xenon, it is around $E_\mathrm{thr} \sim 1$~keV.} has prompted an impressive flourishing of theoretical activity in the search for new physics, inspiring an exhaustive number of phenomenological analyses.

Strikingly, the increase in target size and sensitivity of current DD detectors has opened a whole new window for the observation of the scattering of solar neutrinos. 
While constituting an irreducible background for DM searches, through the so-called $\nu-$floor~\cite{Billard:2013qya} or fog~\cite{OHare:2021utq}, sizable fluxes of solar neutrinos also provide new opportunities for probing neutrino properties~\cite{AtzoriCorona:2022jeb,deGouvea:2021ymm,Giunti:2023yha} and new physics connected to the neutrino sector~\cite{Cerdeno:2016sfi,Dutta:2017nht,Gelmini:2018gqa, Essig:2018tss, Amaral:2020tga, Dutta:2020che,Amaral:2021rzw,Munoz:2021sad}.

Neutrinos coming from the Sun can induce elastic neutrino-electron scattering (E$\nu$ES) and coherent elastic neutrino-nucleus scattering (CE$\nu$NS) events in DM DD experiments, both having a strong degeneracy with the corresponding DM-electron~\cite{Essig:2018tss,Wyenberg:2018eyv} and DM-nucleus signals~\cite{AristizabalSierra:2021kht}.
Given the current detector technology, for E$\nu$ES (CE$\nu$NS) only the $pp$ and $^{7}$Be ($^8$B) components of the total solar neutrino flux contribute significantly to the detectable event rates. However, since the CE$\nu$NS rate is suppressed at current detectors due to threshold limitations, solar neutrinos at DM DD experiments are currently detected mainly through E$\nu$ES.

In this work we are interested in determining the potential of current DM DD experiments, namely XENONnT, LZ and PandaX-4T in probing new physics in the neutrino sector through E$\nu$ES induced by solar neutrinos. Analyses along this line have been done previously to investigate neutrino properties and BSM interactions. These include, for instance, neutrino electromagnetic properties such as effective~\cite{AtzoriCorona:2022jeb,Giunti:2023yha} and fundamental~\cite{A:2022acy} neutrino magnetic moments. 
Future data from e.g. a DARWIN-type experiment will allow for precision searches of electroweak and oscillation parameters as detailed in Ref.~\cite{DARWIN:2020bnc}, while the possibility of probing flavor-dependent radiative corrections to CE$\nu$NS has been explored in Ref.~\cite{Mishra:2023jlq}. 
Neutrino nonstandard interactions (NSI) have been addressed in Ref.~\cite{Amaral:2023tbs} using existing data (XENONnT and LZ) and providing forecasts for DARWIN, complementing the NSI landscape probed by CE$\nu$NS experiments and oscillation searches. 

Going one step beyond effective NSI, our goal is to explore new neutrino interactions mediated by light vector bosons, that can arise in several motivated $U(1)'$~\cite{Langacker:2008yv} extensions of the SM. We will focus on light vector mediators, that are expected to generate
spectral distortions, through an increase of the E$\nu$ES differential cross section, for sufficiently small mediator masses and low recoil energies. DM DD experiments, characterised by low-energy thresholds, are suitable facilities to probe such spectral features~\cite{Boehm:2020ltd,AristizabalSierra:2020edu,Bloch:2020uzh}. Note that in the limit of effective interactions, or equivalently, heavy mediators (with masses much larger that the typical momentum transfer of the process) the NSI formalism used in Ref.~\cite{Amaral:2023tbs} can be directly related to the $U(1)'$ models that we will consider, see, e.g., Ref.~\cite{Coloma:2020gfv}.

We will focus on several BSM models with an extended gauge sector, namely an additional  $U(1)'$ symmetry. These will include: the anomaly-free $B-L$ model~\cite{Langacker:2008yv,Mohapatra:2014yla,Okada:2018ktp}, $B$ being the baryon number and $L$ the total lepton number; lepton flavor-dependent interactions like the $L_e - L_\mu$, $L_\mu - L_\tau$ and $L_e - L_\tau$~\cite{Foot:1990mn,Foot:1990uf,Foot:1992ui,He:1990pn} $U(1)'$ models, also free of quantum anomalies; the $B-2L_e - L_\mu$ and $B-2L_e - L_\tau$ models~\cite{delaVega:2021wpx,AtzoriCorona:2022moj}, where leptons have generation-dependent charges under the $U(1)'$ symmetry; the $B-3L_e$~\cite{Han:2019zkz,delaVega:2021wpx,Coloma:2020gfv,Heeck:2018nzc,AtzoriCorona:2022moj,Barman:2021yaz}, $B-3L_\mu$ and $B-3L_\tau$ scenarios \cite{Ma:1997nq,Ma:1998dp,Ma:1998dr,Ma:1998zg,Chang:2000xy,Heeck:2018nzc,delaVega:2021wpx,Coloma:2020gfv,Barman:2021yaz} and $L_e+2L_\mu+2L_\tau$ \cite{Coloma:2020gfv}.

There is a number of previous works involving light-mediator analyses of DM DD data: Ref.~\cite{A:2022acy} analyzed general light mediators exploiting recent data from the LZ~\cite{LZ:2022lsv} and XENONnT~\cite{XENON:2022ltv} experiments, while Ref.~\cite{Khan:2020csx} performed a similar study employing recent data by the PandaX-II~\cite{PandaX-II:2020udv} experiment. Reference \cite{Schwemberger:2022fjl} set constraints on general neutrino interactions using SENSEI, Edelweiss, and SuperCDMS data and as well as projections 
for future experiments. 
These studies complement bounds inferred using CE$\nu$NS data by COHERENT~\cite{DeRomeri:2022twg} and E$\nu$ES data at Borexino~\cite{Coloma:2022umy}, as well as forecasts for the DUNE near detector~\cite{Melas:2023olz} and FASER~\cite{Asai:2023mzl}. 
Additional studies have focused on anomaly-free $U(1)'$ models. For instance, Ref.~\cite{Park:2023hsp} explored the implications of XENONnT and Borexino E$\nu$ES data on the $B-L$ model, while Refs.~\cite{Amaral:2020tga,Amaral:2021rzw} focused on projections for the $L_\mu-L_\tau$ model at DARWIN. More recently, Ref.~\cite{Majumdar:2021vdw} explored general light mediators as well as $B-L$ and $L_\mu - L_\tau$ gauge symmetries via both CE$\nu$NS and E$\nu$ES at next generation DM DD experiments. Solar neutrino constraints on light mediators have been set by looking at CE$\nu$NS data from CDEX-10 in Ref.~\cite{Demirci:2023tui}.

Our work adds to these previous studies by means of a detailed analysis of several motivated $U(1)'$ scenarios based on the most recent data from the XENONnT~\cite{XENON:2022ltv}, LZ~\cite{LZ:2022lsv}, and PandaX-4T~\cite{PandaX:2022ood} experiments and by providing sensitivities expected at a future DARWIN-like facility.
Moreover, our present analysis improves upon previous E$\nu$ES-based work on light mediators~\cite{A:2022acy} as described in the following. First, we use real, recent data from the current most sensitive experiments. Besides providing separate bounds for each of them, we also present for the first time a combined analysis. Secondly, we carry out an improved statistical analysis by treating the background components of LZ, XENONnT and PandaX-4T separately, together with their own uncertainty, closely following Ref.~\cite{Giunti:2023yha}. Finally, we perform a comprehensive exploration of various vector mediators that go beyond the simplest $B-L$ model, as already introduced.
Our work is aimed at extending and complementing the results obtained in Ref.~\cite{AtzoriCorona:2022moj}, that addresses bounds on several types of $U(1)'$ models using recent COHERENT data.

The remainder of the work is organized as follows: in Section~\ref{sec:theory} we discuss the effect of additional $U(1)'$ symmetries on the E$\nu$ES process. In Section~\ref{sec:analysis} we detail the analysis procedure for current~(\ref{sec:analysis_curr}) and future~(\ref{sec:analysis_fut}) DM DD experiments. In Section~\ref{sec:res} we present the results of our analyses and compare with other bounds from the literature before closing in Section~\ref{sec:conc}.

\section{Theoretical framework}
\label{sec:theory}
We are interested in determining the potential of current DM DD experiments, to probe new light vector bosons through E$\nu$ES induced by solar neutrinos on atomic electrons in the detectors' targets. In this section we introduce the E$\nu$ES process, and provide its cross section both within the SM and in the framework of BSM extensions with an additional $U(1)'$ symmetry.

\subsection{E$\nu$ES in the Standard Model}
 In the SM, the differential E$\nu$ES cross section, with respect to the electron recoil energy $T_e$, is given by

\begin{align}
\label{eq:xsec_vES_SM}
\begin{split}
\frac{d\sigma_{\nu_{\alpha} \mathcal{A}}}{d T_e}\Big|^\mathrm{SM}=& Z^\mathcal{A}_\mathrm{eff}(T_e) \frac{G_F^2m_e}{2\pi}\left[(g_V + g_A)^2 +(g_V - g_A)^2\left(1-\frac{T_e}{E_\nu}\right)^2 -(g_V^2-g_A^2)\frac{m_e T_e}{E_\nu^2}\right] \, ,
\end{split}
\end{align}
where $G_F$, $E_\nu$ and $m_e$ denote the Fermi constant, the incoming neutrino energy and the electron mass, respectively. The SM vector and axial-vector couplings depend on the flavor $\alpha$ of the incoming neutrino $\nu_\alpha$ and read  $g_V=2 \sin^2 \theta_W - 1/2 + \delta_{{\alpha} e}$, $g_A=-1/2 + \delta_{{\alpha} e}$,  $\sin^2 \theta_W=0.23857$ being the weak mixing angle in the low-energy regime. The term $\delta_{{\alpha} e}$ is due to the fact that in the case of incoming electron neutrinos both neutral and charged currents are relevant, while for the case of $\nu_{\mu}$ and $\nu_{\tau}$ E$\nu$ES acquires contributions from neutral currents only. Assuming that the target electrons are bound in the atomic nuclei $\mathcal{A}$ of the detector material, the factor $Z^\mathcal{A}_\text{eff}(T_e)$ in Eq.~(\ref{eq:xsec_vES_SM}) accounts for the effective number of electrons that can be ionized given an energy dissipation $T_e$, and it is taken from the Hartree-Fock calculations provided in Ref.~\cite{Chen:2016eab}. 

\subsection{E$\nu$ES in models with an extra $U(1)'$ symmetry}
Within the context of a general $U(1)'$ symmetry, with a new vector mediator $Z'$ with mass $m_{Z'}$, the Lagrangian describing neutrino-fermion interactions reads 

\begin{equation}
\mathcal{L}_{Z'} =  g_{Z'} Ζ'_\mu \left({Q^f_{Z'}} \bar{f} \gamma^\mu f + \sum_\alpha {Q_{Z'}^{\nu_\alpha}}  \bar{\nu}_{\alpha,L} \gamma^\mu  \nu_{\alpha,L}\right) + \frac{1}{2} {m_{Z'}^2} {Z'^{\mu} Z'_\mu} \, ,
\end{equation}
where for the case of E$\nu$ES $f=e$, ${g_{Z'}}$ is the new coupling while ${Q^i_{Z'}}$ denote the individual vector charges (see Tab.~\ref{tab:U(1)_charges}). Then, for incoming neutrinos of flavor $\alpha$, the new contribution to E$\nu$ES can be obtained from the SM expression of the cross section given in Eq.~(\ref{eq:xsec_vES_SM}) by changing
\begin{equation}
    g_V \to g_V^{\mathrm{SM}} + \frac{({g_{Z'}})^2 Q^e_{Z'} Q^{\nu_{\alpha}}_{Z'}}{\sqrt{2}G_F (2 m_e T_e + m_{Z'}^2)} \, .
    \label{eq:direct_coupling}
\end{equation}
The above expression is valid in both low- and high-mass regimes, noting that the typical momentum transfer involved in DM DD experiments is $|\mathbf{q}| \approx \sqrt{2 m_e T_e} \sim \mathcal{O}(100)$~keV.

In this paper we consider the following models: $B-L$,  $B-3 L_\alpha$, $B-2 L_e - L_{\mu,\tau}$, $L_\alpha - L_\beta$  and $L_e+2L_\mu+2L_\tau$ with the corresponding charges summarized in Tab.~\ref{tab:U(1)_charges}. 
For completeness, let us stress that unlike the $B-L$ case where all neutrino fluxes  contribute equally to new physics cross sections, in lepton flavor-dependent models the new physics contributions come only from the corresponding flavor component of the total solar neutrino flux, weighted through the charges given in Tab.~\ref{tab:U(1)_charges}.  

On the other hand, in the case of the leptophilic model $L_\mu - L_\tau$, since there is no direct coupling to electrons, the new contribution arises at the one-loop level~\cite{AtzoriCorona:2022moj}\footnote{For the sake of completeness let us stress that for the case of CE$\nu$NS there is a plus sign before the second term in Eq.~(\ref{eq:La-Lb_contribution}) because of the different electric charge of protons versus electrons (see Ref.~\cite{Majumdar:2021vdw}).}:  
\begin{equation}
    g_V \to g_V^{\mathrm{SM}} - \frac{\sqrt{2} \alpha_{\mathrm{em}} g_{Z'}^2   (\delta_{\alpha \mu} - \delta_{\alpha \tau})}{\pi G_F (2m_e T_e + m_{V}^2)} \epsilon_{\tau \mu}(|\vec{q}|) \, ,
    \label{eq:La-Lb_contribution}
\end{equation}
where $\alpha_\mathrm{em}$ is the fine-structure constant while the couplings $\epsilon_{\tau \mu}$ can be approximated as
\begin{equation}
    \epsilon_{\tau \mu}(|\vec{q}|) = \int_0^1 x (1-x) \ln \left(\frac{m_\tau^2 + x(1-x) |\vec{q}|^2}{m_\mu^2 + x(1-x) |\vec{q}|^2} \right) \, dx  \approx \frac{1}{6}\ln\left(\frac{m_\tau^2}{m_\mu^2}\right)\, .
\end{equation}
Similarly, in the framework of the $B-3 L_{\mu,\tau}$ models the corresponding E$\nu$ES contribution also arises at the
 one-loop level and reads
\begin{equation}
    g_V \to g_V^{\mathrm{SM}} - \frac{\sqrt{2} \alpha_{\mathrm{em}} g_{Z'}^2 Q^\alpha_{Z'} \delta_{\alpha (\mu/\tau)}}{\pi G_F (2m_e T_e + m_{V}^2)}\tilde{\epsilon}(|\vec{q}|)  \, ,
\end{equation}
where this time $\tilde{\epsilon}$ is given by
\begin{equation}
    \tilde{\epsilon}(|\vec{q}|)   \approx  \frac{1}{6}\sum_f Q^f_{Z'}e_f\ln\left(\frac{m_f^2}{\Lambda^2}\right) \, ,
\label{eq:eps_B-3Lx}
\end{equation}
where $f$ runs over $\mu/\tau$ and all quarks. The couplings $Q^f_{Z'}$ are given in Tab.~\ref{tab:U(1)_charges} and $e_f$ denotes the electric charge of fermion $f$. Finally, $\Lambda$ is the renormalization scale. While an accurate determination would require the full RGE running that is out of the scope of this work, for the sake of simplicity we fix $\Lambda = m_\mu$ ($m_\tau$) in the case of $B-3L_\mu$ ($B-3L_\tau$). We have checked that by choosing a different scale, e.g. $\Lambda = 10^6$~GeV, $|\tilde{\epsilon}|$ would change by a factor of around 4 thus resulting in a bound a factor of 2 stronger. 

\begin{table}[t!]
    \centering
    \begin{tabular}{|c|cc|ccc|}
\hline
         Model & ~~$Q_{Z'}^u~~$ & ~~$Q_{Z'}^d~~$ & ~~$Q_{Z'}^{e/\nu_e}$~ & ~$Q_{Z'}^{\mu/\nu_\mu}$~ & ~$Q_{Z'}^{\tau/\nu_\tau}$~~\\ \hline
         $B-L$ & 1/3 & 1/3 & -1 & -1& -1\\
         \hline
         $B-3L_e$ & 1/3 & 1/3 & -3 & 0 & 0\\
         $B-3L_\mu$ & 1/3 & 1/3 & 0 & -3 & 0\\
         $B-3L_\tau$ & 1/3 & 1/3 & 0 & 0 & -3\\
         \hline
         $B-2L_e - L_\mu$ & 1/3 & 1/3 & -2 & -1 & 0\\
         $B-2L_e - L_\tau$ & 1/3& 1/3 & -2 & 0 & -1\\
         \hline
         $L_e - L_\mu$ & 0 & 0 & 1 & -1 & 0\\         
         $L_e - L_\tau$ & 0 & 0 & 1 & 0 & -1\\
         $L_\mu - L_\tau$ & 0 & 0 & 0 & 1 & -1\\
         \hline
         ~~$L_e+2L_\mu+2L_\tau$~~ & 0 & 0 & 1 & 2 & 2\\
         \hline
    \end{tabular}
    \caption{\label{tab:U(1)_charges} Individual charges in the $U(1)'$ models considered in this work.}
\end{table}

\section{Data analysis}
\label{sec:analysis}

In this section, we present the analysis details for the experiments under consideration. Specifically, we discuss the analysis of current experiments (XENONnT, LZ and PandaX-4T) in subsection~\ref{sec:analysis_curr} and the sensitivity for a next-generation detector, that we model following the DARWIN proposal~\cite{DARWIN:2020bnc}, in subsection~\ref{sec:analysis_fut}. The analysis procedure follows very closely the one described in Ref.~\cite{Giunti:2023yha}. 

\subsection{Current experiments}
\label{sec:analysis_curr}
Regarding current experiments we analyze data from LZ~\cite{LZ:2022lsv}, XENONnT~\cite{XENON:2022ltv} and PandaX-4T~\cite{PandaX:2022ood}. 
The current energy threshold for E$\nu$ES in these experiments is set at $E_\mathrm{thr} = 1$~keV, although due to the detector efficiency the actual flux becomes sizable at $\sim 3$ keV.
For each experiment the number of events due to elastic scattering of solar neutrinos is obtained through the following expression
\begin{equation}
   R^\mathrm{E\nu ES}_k = \mathcal{N} ~\int_{T_e^k}^{T_e^{k+1}} dT_e ~ A(T_e)   ~\int_0^{ T_e^{'\mathrm{max}}} dT_e'~ R(T_e,T_e')~
   \sum_{i=pp,^{7}\text{Be}}\int_{E_{\nu}^{\text{min}}}^{E_{\nu,i}^{\text{max}}} dE_\nu ~\sum_{\alpha} ~\Phi_{\nu_{\alpha}}^{i}(E_\nu)~ \frac{d\sigma_{\nu_{\alpha} \mathcal{A}}}{dT_e'}\,.
\label{eq:r_eves}
\end{equation}
Here, $T_e$ and $T_e'$ are the reconstructed and true electron recoil energies, respectively. The minimal neutrino energy necessary to produce an electron recoil of $T_e'$ is given by $E_\nu^{\text{min}} = (T_e' + \sqrt{2m_eT_e'+T_e'^2})/2$, while the maximal neutrino energy $E_{\nu,i}^{\text{max}}$ is the energy endpoint of the production process inside the Sun, indicated with the index $i$ ($i=pp,^7$Be). The recoil energy upper limit is given by kinematics: $T_e^{'\mathrm{max}} = 2 E_\nu^2 /(2 E_\nu + m_e)$. The E$\nu$ES cross section for a given neutrino flavor $\nu_{\alpha}$ is given by $d\sigma_{\nu_{\alpha} \mathcal{A}} / dT_e'$ and has been discussed in Section~\ref{sec:theory}. The fluxes $\Phi_{\nu_{\alpha}}^{i}(E_\nu)$ are given by 

\begin{equation}
\Phi_{\nu_{e}}^{i}
=
\Phi_{\nu_{e}}^{i\,\odot} P_{ee},
\quad
\Phi_{\nu_{\mu}}^{i}
=
\Phi_{\nu_{e}}^{i\,\odot} \left( 1 - P_{ee} \right) \cos^2\vartheta_{23},
\quad
\Phi_{\nu_{\tau}}^{i}
=
\Phi_{\nu_{e}}^{i\,\odot} \left( 1 - P_{ee} \right) \sin^2\vartheta_{23},
\label{eq:solflux}
\end{equation}
where
$\Phi_{\nu_{e}}^{i\,\odot}$ are the fluxes of $\nu_{e}$ produced
by thermonuclear reactions in the interior of the Sun,
with $i = pp, \, ^{7}\text{Be}$, etc. indicating the production reaction. For those, we employ the spectra from Refs.~\cite{bahcall_web,Bahcall:1987jc,Bahcall:1994cf,Bahcall:1996qv} with the normalizations for the high metallicity model taken from Ref.~\cite{Villante:2020adi}.
As anticipated, the main contributions relevant for the signal at the experiments under consideration originate from $pp$ and $^{7}$Be neutrinos. Moreover, $P_{ee}$ is the $\nu_{e}$-survival probability at the detector on Earth, accounting for neutrino oscillations.
As can be seen, the fluxes of $\nu_{\mu}$ and $\nu_{\tau}$
depend on the mixing angle $\vartheta_{23}$,
which is close to maximal mixing ($\vartheta_{23} \sim \pi/4$)~\cite{deSalas:2020pgw,Esteban:2020cvm,Capozzi:2021fjo}.
For simplicity, we consider
$\sin^2\vartheta_{23}=0.5$,
which implies
$\Phi_{\nu_{\mu}}^{i}=\Phi_{\nu_{\tau}}^{i}$. 

Next, $R(T_e,T_e')$ and $A(T_e)$ are the detector resolution and efficiency functions which are different for all experiments. We use the detector efficiencies given in Refs.~\cite{LZ:2023poo,XENON:2022ltv,PandaX:2022ood}. For the energy resolution at LZ we use the same function that has been used in Refs.~\cite{AtzoriCorona:2022jeb,A:2022acy}. In the case of XENONnT we use the resolution function of Ref.~\cite{XENON:2020rca} and for PandaX-4T we use the one from Ref.~\cite{PandaX:2022ood}. 
Finally, $\mathcal{N} = \mathcal{E} N_T $ is a normalization constant which takes into account the exposure $\mathcal{E}=\{1.16, 0.90, 0.63\}~\mathrm{ton \times year}$ for XENONnT, LZ and PandaX-4T, respectively, while $N_T=N_A/M(\mathrm{Xe})$ denotes the number of target nuclei per $\mathrm{ton}$ of  detector material, $N_A$ and $M(\mathrm{Xe})$ being the Avogadro number and the molar mass of xenon. 

The overall predicted number of events in a given energy-bin $k$ for an experiment $X$ is given by 
\begin{equation}
    R^X_k = R^\mathrm{E\nu ES}_k + \sum_i R^i_k\,,
\label{eq:pred_n_evs}
\end{equation}
where $R^\mathrm{E\nu ES}_k$ is the E$\nu$ES contribution, Eq.~\eqref{eq:r_eves}, while $R^i_k$ are the remaining background components of each experiment, extracted from Refs.~\cite{LZ:2022ysc,XENON:2022ltv,PandaX:2022ood} for LZ, XENONnT and PandaX-4T, respectively.
The total number of predicted events has to be compared with the data $D^X$ collected in each experiment. In our analyses we use the data from Fig.~6 of Ref.~\cite{LZ:2022lsv} for LZ and the data from Fig.~3 of Ref.~\cite{PandaX:2022ood} for PandaX-4T. In the case of XENONnT we use the data from Fig.~4 (5) of Ref.~\cite{XENON:2022ltv}, for recoil energies above (below) 30~keV. 

Due to the low statistics in some bins, for LZ and PandaX-4T we use the Poissonian least-squares function
\begin{equation}
    \chi^2_X = \min_{\vec{\alpha},\vec{\beta}} \left\{2\left(\sum_k R^X_k - D^X_k + D^X_k~\ln \left(D^X_k/R^X_k\right)\right) + \sum_i (\alpha_i/\sigma_{\alpha_i})^2 + \sum_i (\beta_i/\sigma_{\beta_i})^2\right\}\,,
\end{equation}
where $\alpha_i$ are normalization constants multiplied to each single background component in Eq.~\eqref{eq:pred_n_evs}, penalized by the uncertainties $\sigma_{\alpha_i}$.
These uncertainties are taken from Tab.~VI of Ref.~\cite{LZ:2022ysc} for LZ and Tab.~I of Ref.~\cite{PandaX:2022ood} for PandaX-4T.
As indicated in the references, some of these nuisance parameters are left to vary freely in the analysis.
Also included are the uncertainty coefficients  $\beta_i$ of the solar neutrino fluxes, with uncertainties $\sigma_{\beta_i}$ taken from Ref.~\cite{Villante:2020adi}. 

In the case of XENONnT data, we use instead
\begin{equation}
    \chi^2_{\text{XENONnT}} = \min_{\vec{\alpha},\vec{\beta}} \left\{\sum_k\left(\frac{R^{\text{XENONnT}}_k - D^{\text{XENONnT}}_k}{\sigma_k}\right)^2 + \sum_i (\alpha_i/\sigma_{\alpha_i})^2 + \sum_i (\beta_i/\sigma_{\beta_i})^2\right\}\,,
\end{equation}
where the uncertainties in each bin $\sigma_k$ are extracted from Ref.~\cite{XENON:2022ltv}. The remaining components are equivalent to the corresponding ones for LZ and PandaX-4T. 

We also perform a combined analysis of all three experiments by correlating the uncertainties regarding the neutrino flux among the experiments. In addition, background components which are common to at least two of the three experiments are also correlated. 

\subsection{DARWIN sensitivity}
\label{sec:analysis_fut}

The calculation of the event rate at a future experiment like DARWIN is essentially the same as for the current experiments, given in Eqs.~\eqref{eq:r_eves} and~\eqref{eq:pred_n_evs}. The only difference is that we also include the contributions from solar N, O and $pep$ neutrinos. It should be noted, however, that their contribution is mostly negligible in comparison with some of the background contributions, as shown in Fig.~1 of Ref.~\cite{DARWIN:2020bnc}. We include them, nevertheless, since we use the full spectrum as shown in Ref.~\cite{DARWIN:2020bnc}. 

The individual background components relevant for DARWIN are given in Ref.~\cite{DARWIN:2020bnc}, and need to be normalized to the considered exposure. We use the same resolution function and detector efficiency as for XENONnT assuming that the efficiency remains constant for $T_e>T_{e,\text{max}}^{\text{XENONnT}}$. With these assumptions, we are able to reproduce the E$\nu$ES spectra for all five neutrino species in Fig.~1 of Ref.~\cite{DARWIN:2020bnc}, which justifies the choices of efficiency and resolution functions. We compute the sensitivity considering an exposure of $30~\mathrm{ton \times years}$ and $300~\mathrm{ton \times years}$. 

When generating the mock data, always compatible with the SM  expectation, we use 51 logarithmically-spaced bins between 1 and 1500~keV recoil energy. Note that the spectrum at higher energies ($\gtrsim 700$ keV) is not sensitive to any BSM effect considered in this paper, because the E$\nu$ES rate is much smaller than some of the background rates. Indeed, the main E$\nu$ES rate contributions from $pp$ and $^{7}$Be neutrinos become irrelevant at $T_e \sim 250$ keV and $\sim 700$ keV, respectively. We still use the full spectrum up to 1500~keV, since the inclusion of events at high energies can help to control the effect of background uncertainties.

\section{Results}
\label{sec:res}

In this section we present the bounds obtained for all models discussed in Section~\ref{sec:theory}. In all cases we compare our bounds with previous ones obtained in the literature, in particular with the ones included in the DarkCast package~\cite{Ilten:2018crw}, the ones obtained from COHERENT data~\cite{AtzoriCorona:2022moj,Melas:2023olz} and from neutrino oscillation experiments~\cite{Coloma:2020gfv}. 
DarkCast implements constraints from several classes of experiments\footnote{Note that even though not all DarkCast bounds are at the same confidence level, small differences in confidence level are basically invisible on the scales of our plots.}:
beam dump
(E141~\cite{Riordan:1987aw},
E137~\cite{Bjorken:1988as},
E774~\cite{Bross:1989mp},
KEK~\cite{Konaka:1986cb},
Orsay~\cite{Davier:1989wz,Bjorken:2009mm,Andreas:2012mt},
$\nu$-CAL~I~\cite{Blumlein:1990ay,Blumlein:1991xh,Blumlein:2011mv,Blumlein:2013cua},
CHARM~\cite{CHARM:1985anb,Gninenko:2012eq},
NOMAD~\cite{NOMAD:2001eyx},
and
PS191~\cite{Bernardi:1985ny,Gninenko:2011uv}), 
fixed target
(A1~\cite{Merkel:2014avp} and APEX~\cite{APEX:2011dww}),
colliders
(BaBar~\cite{BaBar:2014zli},
KLOE~\cite{KLOE-2:2011hhj,KLOE-2:2016ydq},
LHCb~\cite{LHCb:2017trq}), rare-meson decay (NA48/2~\cite{NA482:2015wmo}) experiments, neutrino scattering experiments (TEXONO, CHARM-II and BOREXINO~\cite{Bauer:2018onh,TEXONO:2009knm,Harnik:2012ni,Bellini:2011rx,CHARM-II:1993phx}), and
searches for dark photons in  
NA64~\cite{NA64:2019auh,NA64:2022yly,NA64:2023wbi,Andreev:2024sgn} and BaBar~\cite{BaBar:2017tiz}.\\

\subsection*{The $B-L$ model}

\begin{figure}[t]
\centering    
\includegraphics[width=0.95\textwidth]{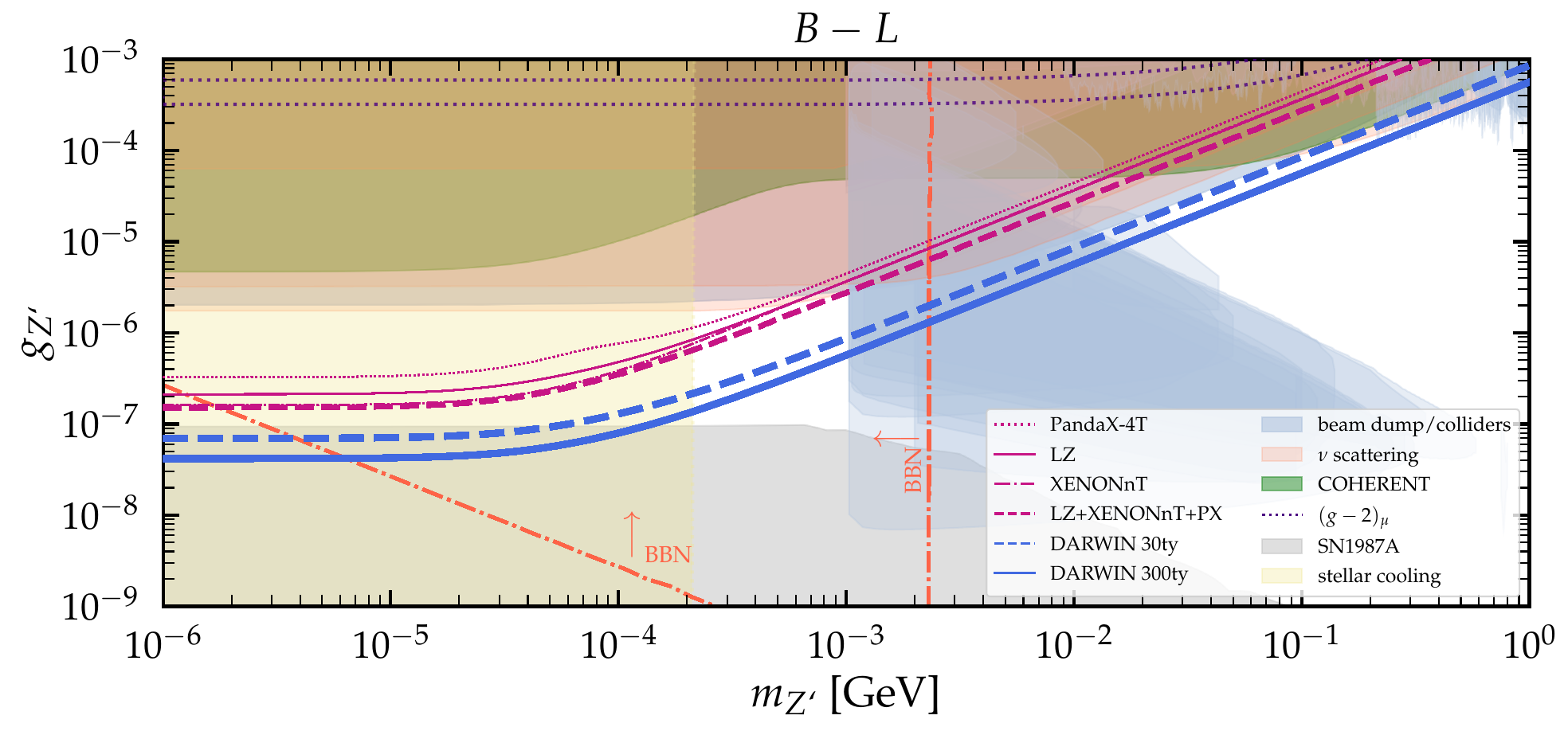}
\caption{$90 \%$ C.L. bounds from our analyses of: LZ (magenta solid), PandaX-4T (magenta dotted) and XENONnT data (magenta dashed-dotted), a combined analysis of current data (magenta dashed) and expected sensitivity at DARWIN with an exposure of 30  (blue dashed) or 300~$\mathrm{ton \times years}$ (blue solid) for the $B-L$ model, in comparison with other existing bounds. }
\label{fig:B-L}
\end{figure}

The $90\%$ C.L. exclusion limits in the $m_{Z'} - g_{Z'}$ plane for the $B-L$ model are shown in Fig.~\ref{fig:B-L}. Our present results for XENONnT and LZ are in excellent agreement with a previous analysis performed in Ref.~\cite{A:2022acy}, though here the analysis is improved following the method of Ref.~\cite{Giunti:2023yha}, as explained previously. The thin magenta lines correspond to the analyses of PandaX-4T (dotted), LZ (solid) and XENONnT (dashed-dotted) data, while the thick magenta dashed line is the result from our combined analysis\footnote{In order to not overcrowd the figures we will show only the combined result in subsequent figures.}. At large (small) masses the dominating contribution to the bound comes from LZ (XENONnT) data. Note that since we correlated common uncertainties among the experiments, the combined bound is stronger than a simple sum of $\chi^2$ functions. For low mediator masses ($\lesssim 0.1$ MeV), our combined limit saturates at $g_{Z'} \sim 1.5\times 10^{-7}$.

As can be seen in Fig.~\ref{fig:B-L}, current DM DD experiments produce slightly weaker (stronger) bounds for large (small) masses than other experiments measuring E$\nu$ES~\cite{Bauer:2018onh,TEXONO:2009knm,Harnik:2012ni,Bellini:2011rx,CHARM-II:1993phx}, like TEXONO, CHARM-II and BOREXINO (``$\nu$ scattering", depicted as coral-shaded areas). On the other hand, DM DD bounds do improve upon CE$\nu$NS constraints\footnote{Note that the depicted constraints actually include CE$\nu$NS+E$\nu$ES events in the analysis of COHERENT-CsI data (see Refs.~\cite{DeRomeri:2022twg,Melas:2023olz} for details).} from the COHERENT experiment~\cite{Melas:2023olz} (green area), by about two orders of magnitude at $m_{Z'} \lesssim 0.01$ MeV.
Among beam-dump experiments, usually covering masses $m_{Z'} \gtrsim 1$ MeV, let us highlight NA64~\cite{NA64:2022yly}, whose bound applies to lighter mediators as well and is comparable to the limits set by $\nu$-scattering experiments.
It should be noted that current bounds from DM experiments are never dominating in comparison to other probes for large mediator masses, i.e., $m_{Z'} \gtrsim 10$ MeV. However, DARWIN will provide the dominating bounds in the region $0.02$ GeV $\lesssim m_{Z'} \lesssim 0.4$ GeV, between the blue- and coral-shaded regions corresponding to beam-dump and accelerator experiments and scattering experiments, respectively. The purple dotted lines indicate the region of parameter space that could explain the $(g-2)_\mu$ anomaly~\cite{Aoyama:2020ynm} and that is therefore now excluded by several experiments. Let us finally note that below the $\sim$ MeV scale strong bounds from big bang nucleosynthesis~\cite{Huang:2017egl} (orange dashed-dotted lines) and stellar cooling~\cite{Li:2023vpv} effects (yellow area) 
apply, although their exact estimation will eventually
depend on the specific model and thermal history in the early Universe. Finally, bounds from SN1987A~\cite{Chang:2016ntp,Croon:2020lrf}  (light gray) constrain the region in the plot for small couplings\footnote{Let us note that the latter constraints might be less stringent
as explained in Refs.~\cite{Caputo:2021rux,Caputo:2022rca}.}.

\subsection*{The $B-2L_e-L_{\mu/\tau}$ models}

\begin{figure}[t]
\begin{subfigure}{0.9\textwidth}
    \includegraphics[width=\textwidth]{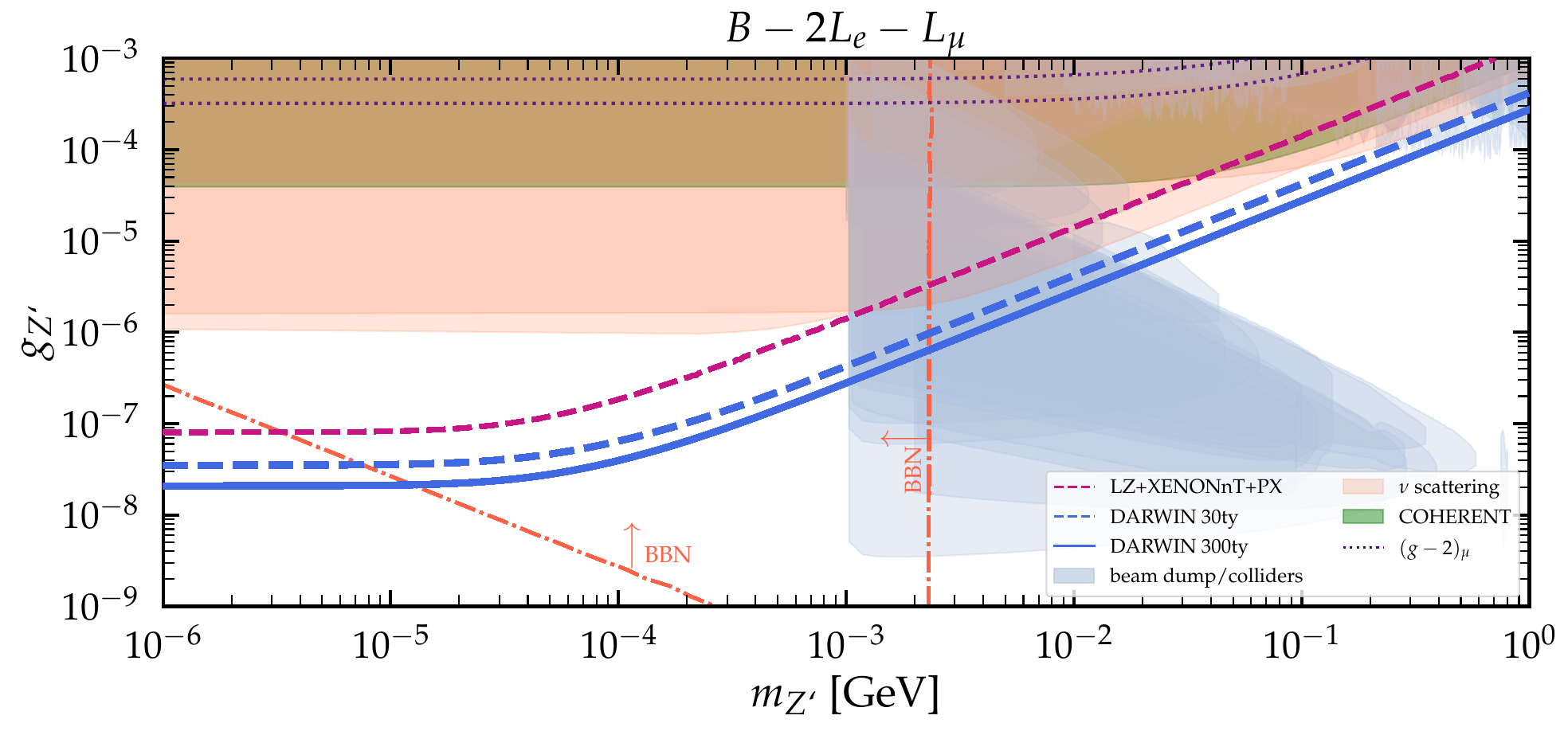}
\end{subfigure}
\hfill
\begin{subfigure}{0.9\textwidth}
    \includegraphics[width=\textwidth]{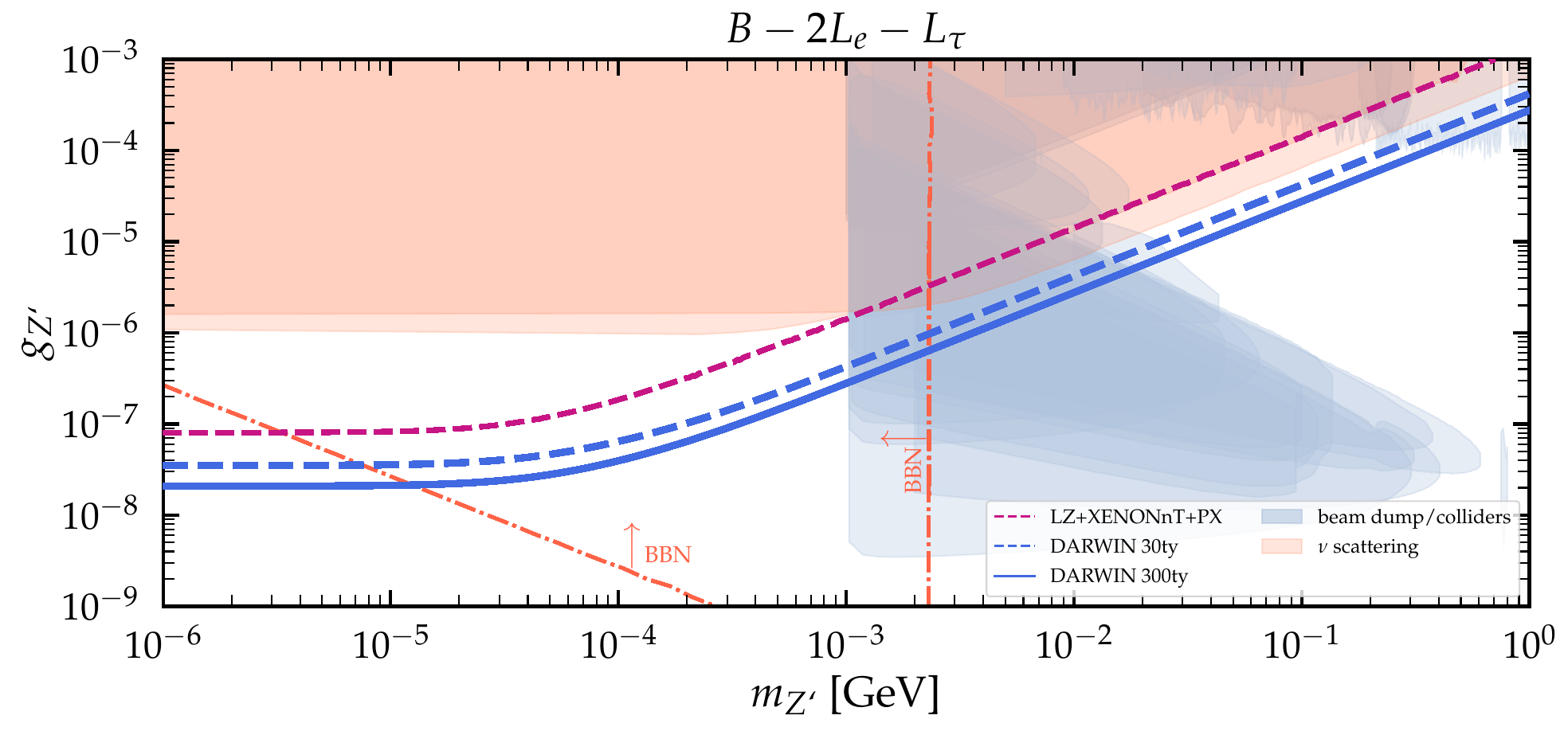}
\end{subfigure}
\caption{$90 \%$ C.L. bounds from our combined analysis of LZ, PandaX-4T and XENONnT data and the sensitivity at DARWIN for the $B-2L_e-L_{\mu}$ (upper panel) and $B-2L_e-L_{\tau}$ (lower panel) models. Also shown are bounds from other experiments for comparison.}
\label{fig:B-2Le-La}
\end{figure}

The upper (lower) panel of Fig.~\ref{fig:B-2Le-La} shows the results for the $B-2L_e-L_{\mu}$ ($B-2L_e-L_{\tau}$)  models~\cite{delaVega:2021wpx,AtzoriCorona:2022moj}. Due to our choice of $\sin^2\theta_{23}=0.5$ the bounds from DM DD experiments are the same for both models. However, the bounds from other experiments can change. Note that although not analyzed in Ref.~\cite{AtzoriCorona:2022moj}, COHERENT data could also be used to place bounds on the $B-2L_e-L_{\tau}$ scenario, however the resulting limit would be slightly weaker than those for $B-2L_e-L_{\mu}$ and definitely weaker than our combined constraint from LZ+PandaX-4T+XENONnT. 
As in the previous case of the $B-L$ model, solar E$\nu$ES data at DM DD experiments will be able to provide the dominating bound in the $0.02$ GeV $\lesssim m_{Z'} \lesssim 0.2$ GeV mass region even after a relatively short exposure time at DARWIN, while the current limit (magenta dashed) is slightly weaker than other probes, but nonetheless comparable to other limits from neutrino scattering at $m_{Z'} \gtrsim 1$ MeV. All in all, at small masses $m_{Z'} \lesssim 1$ MeV the bounds from DM DD experiments are significantly stronger than those from other laboratory probes,  while being surpassed by astrophysical and cosmological observations. 

\begin{figure}[t]
\begin{subfigure}{0.9\textwidth}
    \includegraphics[width=\textwidth]{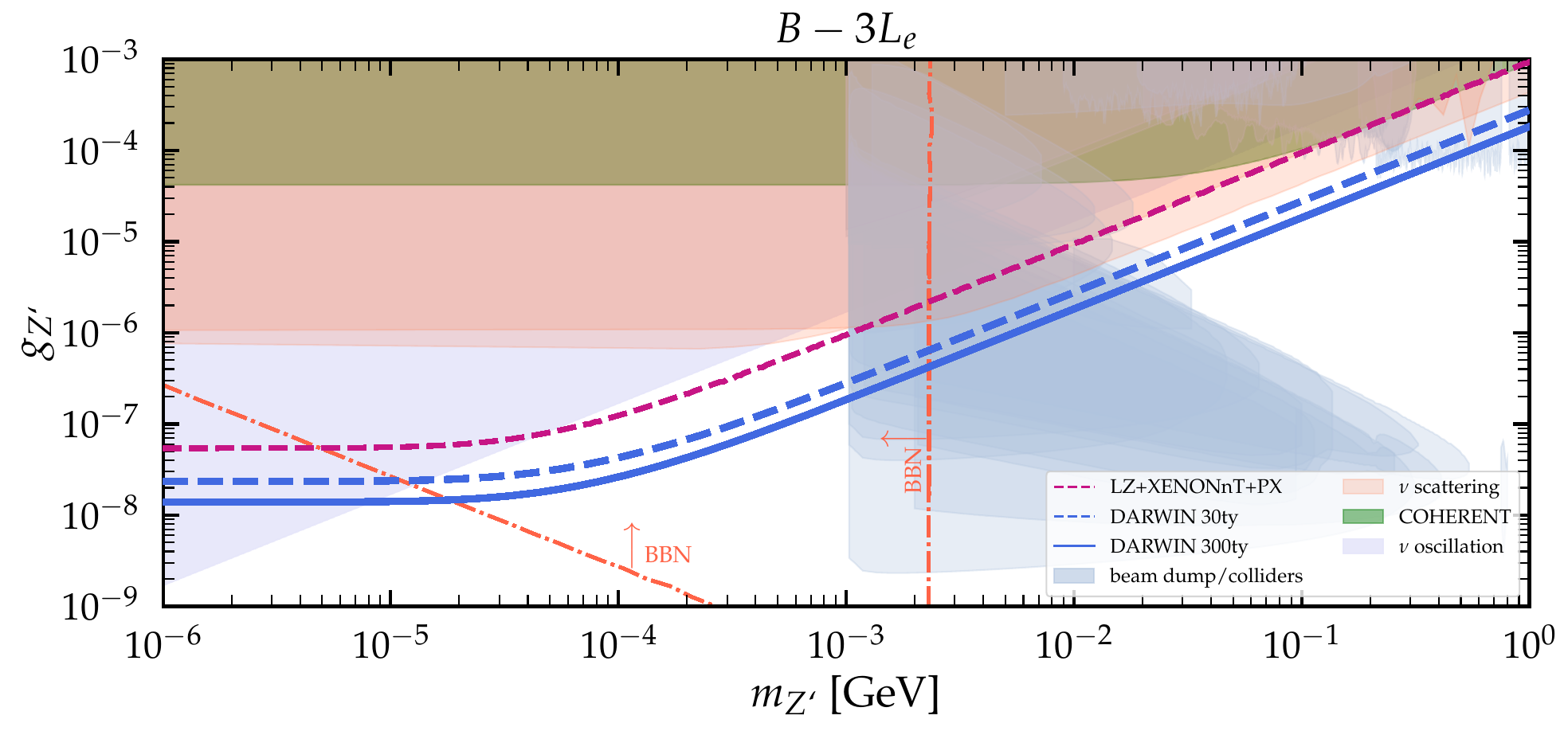}
\end{subfigure}
\hfill
\begin{subfigure}{0.9\textwidth}
    \includegraphics[width=\textwidth]{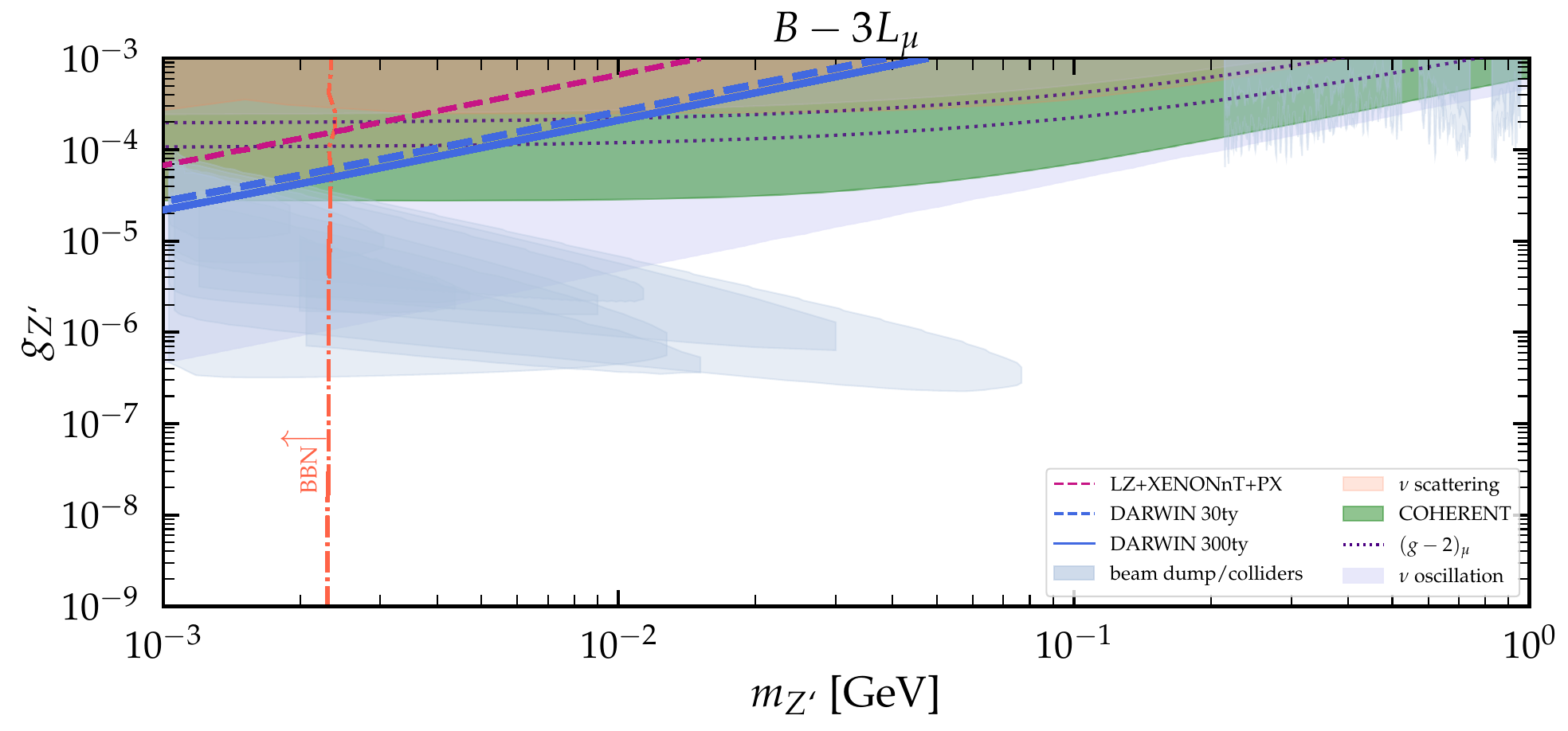}
\end{subfigure}
\hfill
\begin{subfigure}{0.9\textwidth}
    \includegraphics[width=\textwidth]{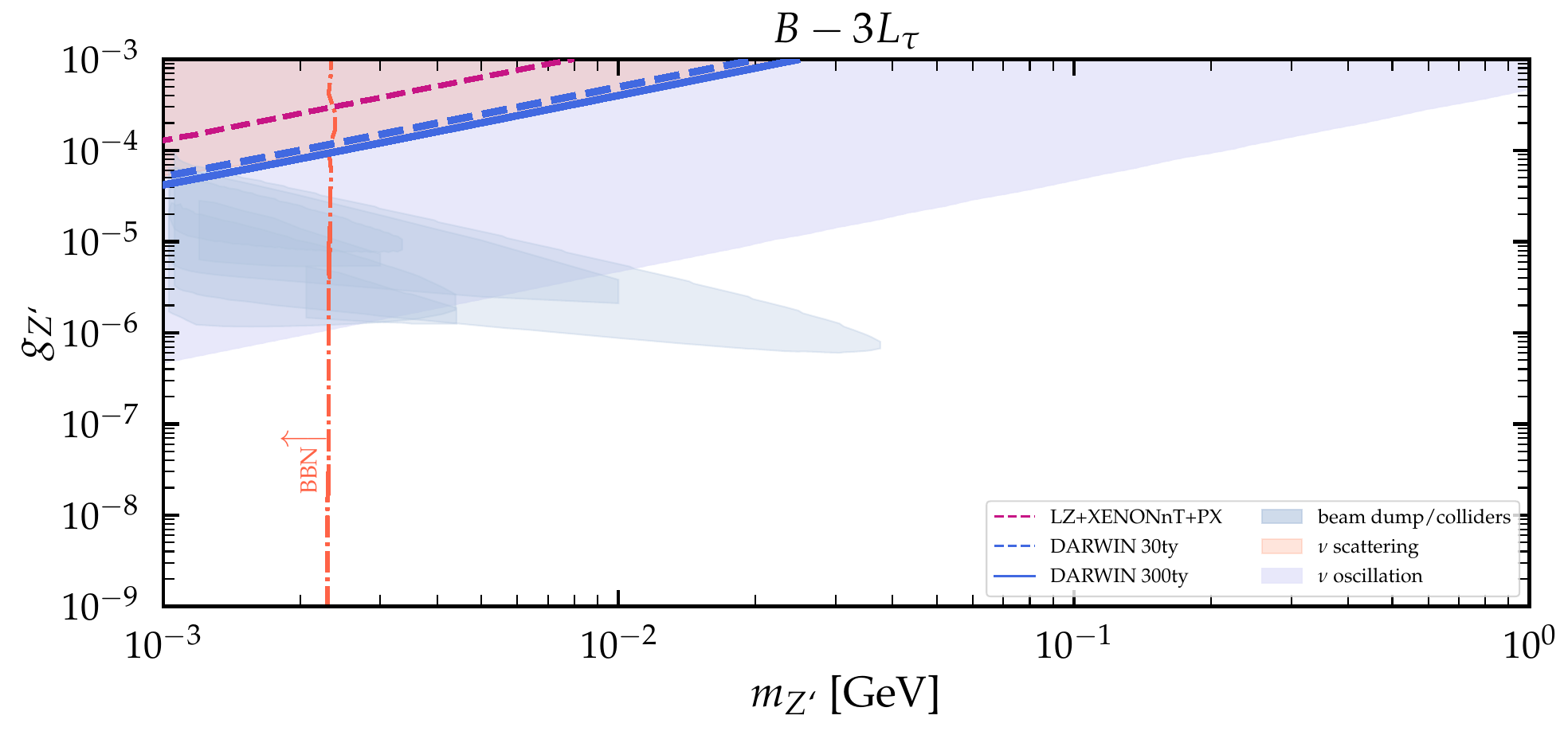}
\end{subfigure}
\caption{The bounds from our combined analysis of LZ, PandaX-4T and XENONnT data and the sensitivity at DARWIN at 90\% C.L. for $B-3L_e$ (first row), $B-3L_\mu$ (second row) and $B-3L_\tau$ (third row). Also shown are bounds from other experiments for comparison.}
\label{fig:B-3La}
\end{figure}

\subsection*{The $B-3L_\alpha$ models}

We next discuss the results obtained for the $B-3L_\alpha$ models. Results are presented in Fig.~\ref{fig:B-3La}. In the case of $B-3L_e$ we find a behavior that resembles previous models. At large mediator masses current DM DD experiments are slightly weaker than other probes, while DARWIN has the prospect to provide the leading bound in the future. At small masses our bounds are the strongest limits obtained using neutrino fluxes in some parts of the parameter space. However, the bound from oscillation experiments~\cite{Coloma:2020gfv} (light lavender) starts to dominate for mediator masses below $\sim \mathcal{O}(10)$~keV.

In the second and third row of Fig.~\ref{fig:B-3La} we show the results for the $B-3L_\mu$ and $B-3L_\tau$ models, respectively. We focus on a reduced range of mediator masses, since for smaller masses a more thorough treatment of the loop-effects in Eq.~\eqref{eq:eps_B-3Lx} might become necessary. As can be seen, our bounds from current DM DD experiments as well as the projected sensitivity for DARWIN seem not to be competitive with oscillation experiments. The reason is that these models affect the E$\nu$ES process only at the loop-level and hence the overall effect on the cross section is very tiny in comparison with the $B-3L_e$ model, for which the interaction occurs instead at tree level.

\subsection*{The $L_\alpha - L_\beta$ models}

\begin{figure}[t]
\begin{subfigure}{0.9\textwidth}
    \includegraphics[width=\textwidth]{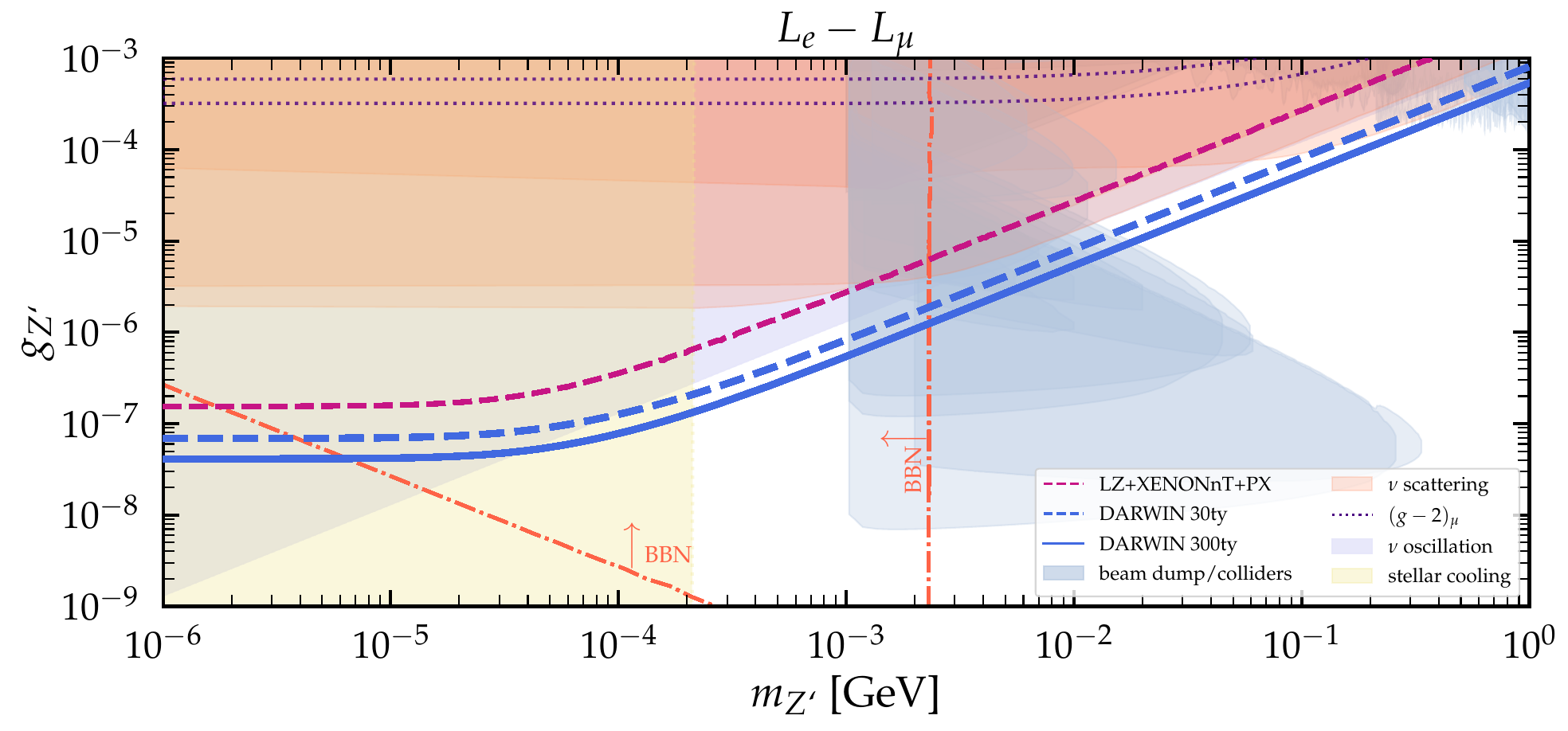}
\end{subfigure}
\hfill
\begin{subfigure}{0.9\textwidth}
    \includegraphics[width=\textwidth]{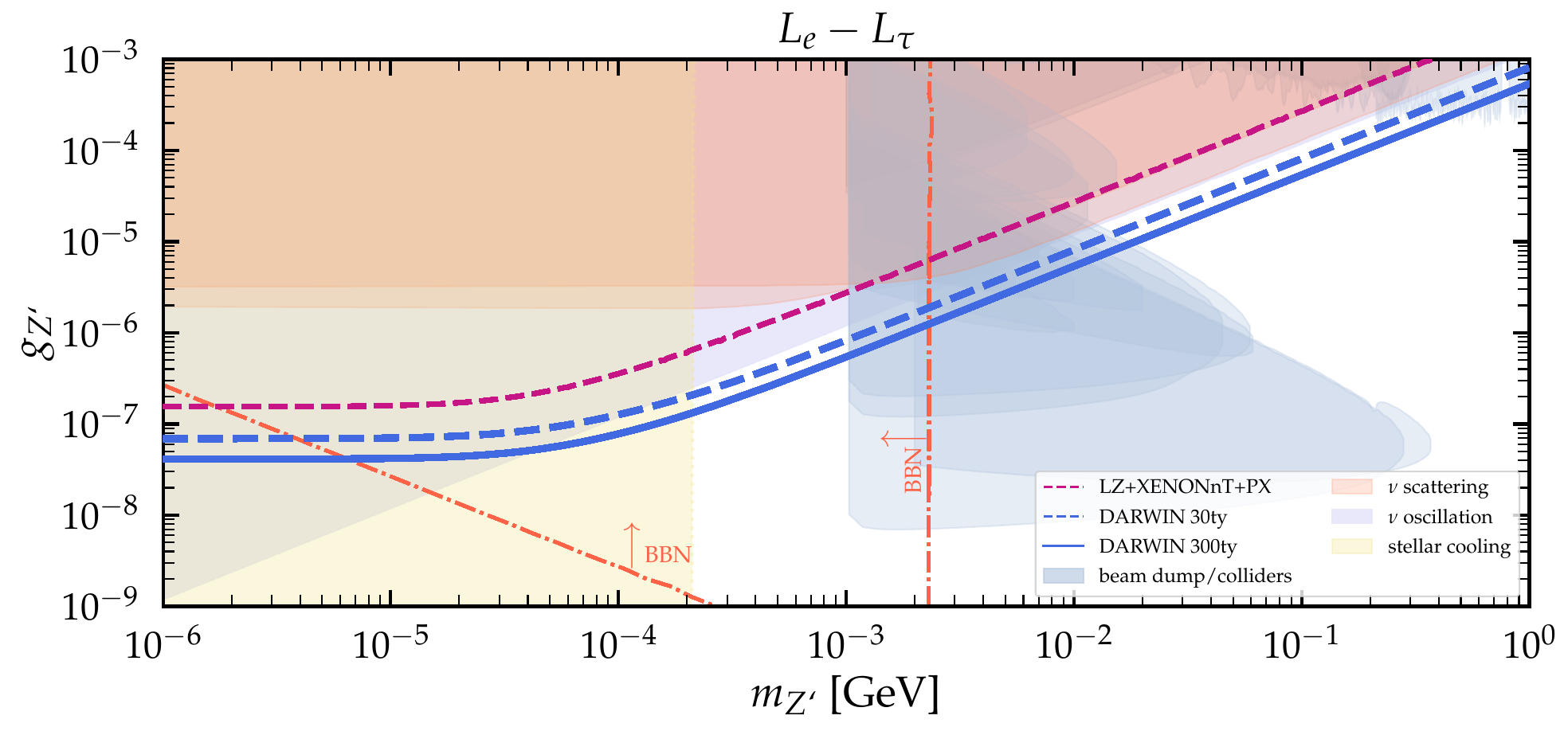}
\end{subfigure}
\hfill
\begin{subfigure}{0.9\textwidth}
    \includegraphics[width=\textwidth]{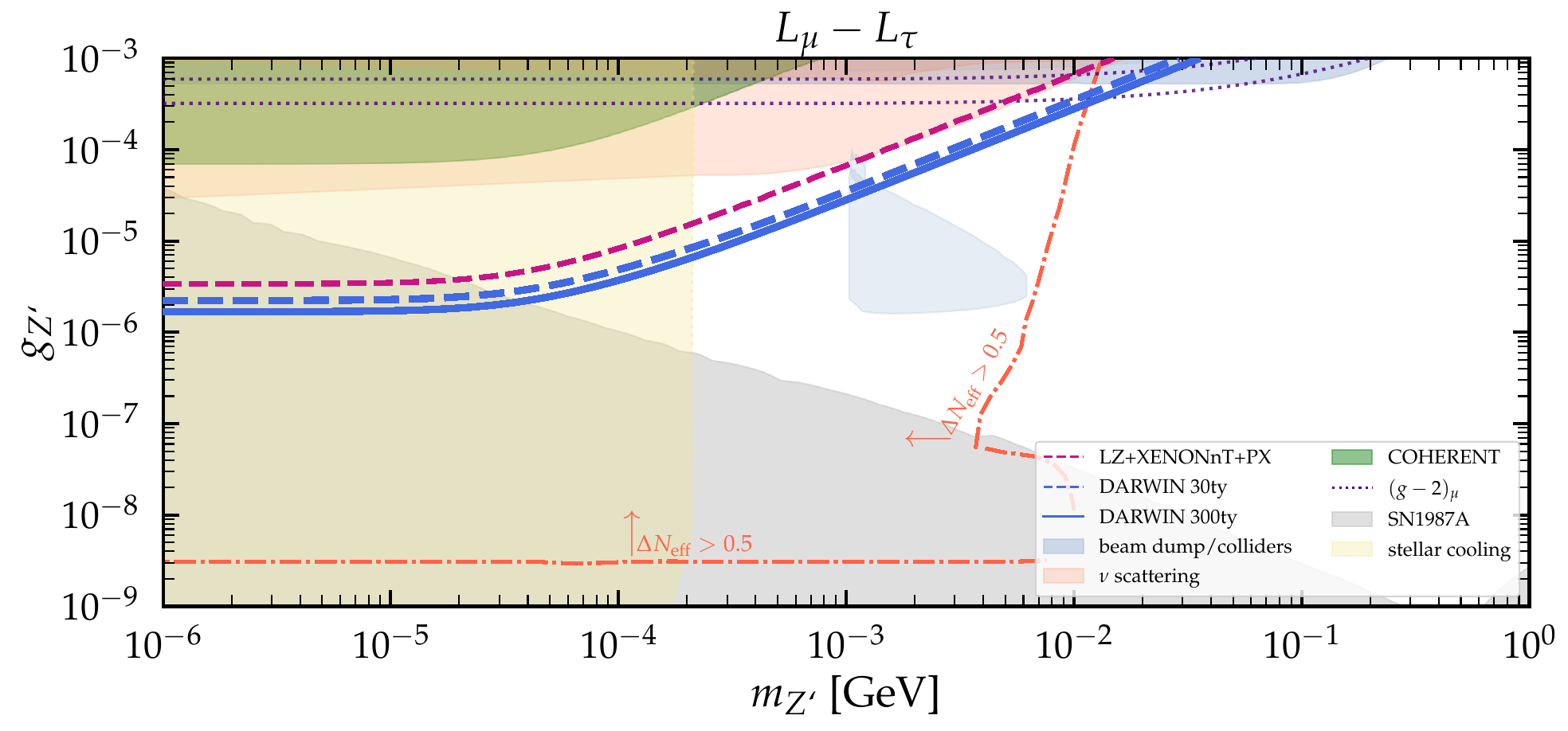}
\end{subfigure}
\caption{The bounds from our combined analysis of LZ, PandaX-4T and XENONnT data and the sensitivity at DARWIN at 90\% C.L. for $L_e-L_\mu$ (first row), $L_e-L_\tau$ (second row) and $L_\mu-L_\tau$ (third row). Also shown are bounds from other experiments for comparison.}
\label{fig:La-Lb}
\end{figure}

In the last two subsections we discuss models that only couple to leptons. We start with the $L_\alpha - L_\beta$ models. Our results are presented in Fig.~\ref{fig:La-Lb}. As in previous cases, we find that DARWIN will be able to provide the dominant bound in some regions of the parameter space, even with relatively small exposure, for $L_e-L_{\mu/\tau}$. For the same symmetry arguments already mentioned above, our bounds and sensitivities are the same for both models. Even though DARWIN will provide the strongest bound at large mediator masses, at  small masses the bounds from oscillation experiments remain very strong as can be seen in Fig.~\ref{fig:La-Lb}.

In the case of the $L_\mu - L_\tau$ model, current bounds from DM DD experiments are of similar strength as those from other E$\nu$ES probes (``$\nu$ scattering") and are expected to become much stronger once DARWIN starts taking data. The current combined limit already rules out $L_\mu - L_\tau$ as en explanation to the $(g-2)_\mu$ anomaly, at $m_{Z'} \lesssim 100$ MeV. Notice also that at $m_{Z'} \gtrsim 100$ MeV the NA64 bound~\cite{Andreev:2024sgn} becomes relevant and excludes couplings $g_{Z'} \gtrsim 5 \times 10^{-3}$. We do not show oscillation bounds for this model, because this scenario was not considered in Ref.~\cite{Coloma:2020gfv}. The COHERENT result, available for the case of $L_\mu - L_\tau$ only, is taken from Ref.~\cite{Melas:2023olz}\footnote{For $g_{Z'}< 10^{-3}$ the COHERENT bound is entirely driven by E$\nu$ES events.}. Let us also mention that bounds for very small mediator masses have been calculated in Ref.~\cite{Alonso-Alvarez:2023tii}. However, for the mass range of interest in this work, those bounds are many orders of magnitude weaker than the ones derived in this work. Finally, a large region of parameter space with small coupling is in conflict with cosmological observations~\cite{Escudero:2019gzq} ($\Delta N_\mathrm{eff}$) and supernova data~\cite{Croon:2020lrf,Fiorillo:2022cdq,Akita:2023iwq}.

\subsection*{The $L_e+2L_\mu+2L_\tau$ model}

\begin{figure}[t]
\begin{subfigure}{0.9\textwidth}
    \includegraphics[width=\textwidth]{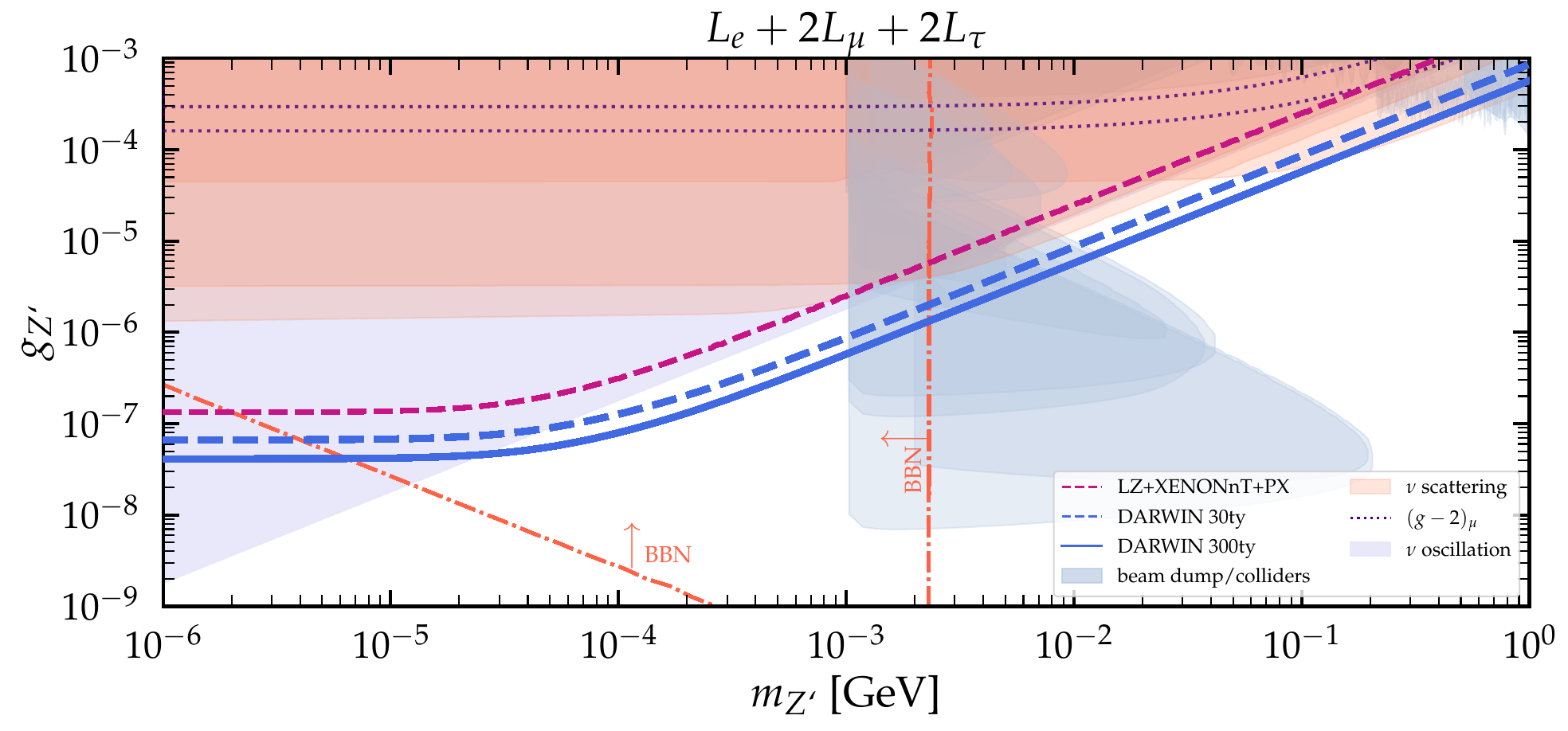}
\end{subfigure}
\caption{The bounds from our combined analysis of LZ, PandaX-4T and XENONnT data and the sensitivity at DARWIN at 90\% C.L. for $L_e+2L_\mu+2L_\tau$. Also shown are bounds from other experiments for comparison.}
\label{fig:Le+2Lm+2Lt}
\end{figure}

The last model that we consider is $L_e+2L_\mu+2L_\tau$. The results are shown in Fig.~\ref{fig:Le+2Lm+2Lt}. As in the case of previous models we find that DM DD experiments currently provide limits comparable to other $\nu-e^-$ scattering experiments but will improve in some parts of the parameter space with future facilities like DARWIN. For low mediator masses, however, DM DD bounds are already much stronger than the ones obtained from other experiments using E$\nu$ES and will be further improved by DARWIN. As in previous scenarios, these bounds probe parts of the parameter space in complementarity to those probed by astrophysical observations or by other terrestrial experiments like beam dump and colliders.

\section{Conclusions}
\label{sec:conc}
The large exposures achieved at recent dark matter direct detection experiments, combined with the very low threshold operation capabilities of recent LXe and future LAr detectors, mark a turning point in solar neutrino detection. 
Motivated by this unique opportunity, we have performed a thorough analysis of compelling $U(1)'$ models, by analyzing current (XENONnT, LZ and PandaX-4T) and future (DARWIN) DM DD experiments via the E$\nu$ES channel. In particular, we have focused on spectral distortions expected in the E$\nu$ES rates that arise in the presence of novel interactions within the anomaly-free $B-L$, $L_\alpha - L_{\beta}$,  $B-2L_e - L_{\mu, \tau}$, $B-3L_\alpha$, and $L_e+2L_\mu +2 L_{\tau}$ models. By means of an improved statistical analysis, in which the various experimental uncertainties are treated separately, we have obtained stringent constraints on the relevant parameter space of the new vector mediator, $(m_{Z'}, g_{Z'})$. We have  presented a combined analysis of  ongoing experiments: XENONnT, LZ and PandaX-4T showing that it leads to slightly improved sensitivities compared to those obtained by analyzing each experiment individually. Specifically, we have shown that for $m_{Z'} \gtrsim 100~\mathrm{MeV}$ current DM DD experiments place competitive constraints, complementing other experimental probes including neutrino oscillation data, beam dump and collider searches. On the other hand, in the low mass regime i.e. for $m_{Z'} \lesssim 1~\mathrm{MeV}$ we have illustrated that the E$\nu$ES channel dominates the constraints among terrestrial experiments. We have further shown that sensitivities achievable at future DM DD experiments like DARWIN ---in view of their large size--- will offer further improvements. We have finally discussed complementarities of our present results with constraints coming from astrophysical observations, especially those obtained from BBN and stellar cooling data.

\section*{Acknowledgments}

We thank Martin K. Hirsch for enlightening discussions, Sergei Gninenko, Laura Molina Bueno and Andrés D. Pérez for useful comments. V.D.R. acknowledges financial support by the CIDEXG/2022/20 grant (project ``D'AMAGAT'') funded by
Generalitat Valenciana and by the Spanish grant PID2020-113775GB-I00 (MCIN/AEI/10.13039/501100011033).
The work of DKP was supported by the Hellenic Foundation for Research and Innovation (H.F.R.I.) under the “3rd Call for H.F.R.I. Research Projects to support Post-Doctoral Researchers” (Project Number: 7036).

\bibliographystyle{utphys}
\bibliography{bibliography}  

\end{document}